\documentclass[11pt]{article}
\usepackage{latexsym,amssymb,amsmath,amsthm,a4wide,setspace,enumerate,color,graphicx,bbm,bm,url,dsfont}
\usepackage[T1]{fontenc}
\usepackage[margin=3cm]{geometry}
\usepackage{natbib}
\usepackage{cmbright}
\usepackage{xcolor}
\usepackage{xcolor-solarized}
\pagecolor{solarized-base2}
\color{solarized-base02}
\usepackage{changepage}

\usepackage{pgf}
\usepackage{graphicx}
\usepackage{epsfig}

\usepackage{tikz}

\usetikzlibrary{arrows,positioning}

 \usepackage{array}
\usepackage{xfrac} 
\usepackage{tablefootnote}
\usepackage[para,online,flushleft]{threeparttable}

\usepackage{fancyhdr}

\fancyheadoffset{0cm}

\fancypagestyle{plain}{%
   \fancyhf{}%
   \fancyhead[R]{\thepage}%
}
\usepackage[framemethod=TikZ]{mdframed}
\usepackage{lipsum}
\mdfdefinestyle{MyFrame}{%
	userdefinedwidth=5cm
	linecolor=black,
	outerlinewidth=1pt,
	roundcorner=10pt,
	innertopmargin=\baselineskip,
	innerbottommargin=\baselineskip,
	innerrightmargin=10pt,
	innerleftmargin=10pt,
	backgroundcolor=white}

\usepackage{afterpage}
\usepackage{titlesec}

  \usepackage{tikz}
  \usetikzlibrary{positioning, calc, shapes.geometric, shapes.multipart,
shapes, arrows.meta, arrows,
decorations.markings, external, trees}

\tikzstyle{Arrow} = [
thick,
decoration={
markings,
mark=at position 1 with {
\arrow[thick]{latex}
}
},
shorten >= 3pt, preaction = {decorate}
]

\newcommand{\slice}[4]{
  \pgfmathparse{0.5*#1+0.5*#2}
  \let\midangle\pgfmathresult

  \draw[thick,fill=black!10] (0,0) -- (#1:1) arc (#1:#2:1) -- cycle;

  \pgfmathparse{min((#2-#1-10)/110*(-0.3),0)}
  \let\temp\pgfmathresult
  \pgfmathparse{max(\temp,-0.5) + 0.8}
  \let\innerpos\pgfmathresult
  \node at (\midangle:\innerpos){#4};
}
\usepackage{xfrac}
  
\usepackage{graphicx}
\usepackage{subcaption}

\usepackage{authblk}
\author[1,2,3]{Anders Huitfeldt}
\author[4,5]{Matthew P. Fox}
\author[4]{Eleanor J. Murray}
\author[1,2]{Asbj\o rn Hr\'{o}bjartsson}
\author[6]{Rhian M. Daniel}
\affil[1]{Centre for Evidence-Based Medicine Odense (CEBMO) and Cochrane Denmark, Department of Clinical Research, University of Southern Denmark, Odense, Denmark
}
\affil[2]{Open Patient data Exploratory Network (OPEN), Odense University Hospital, Odense, Denmark}
\affil[3]{Department of Mathematics, Ecole Polytechnique F\'ed\'erale de Lausanne (EPFL), Switzerland}
\affil[4]{Department of Epidemiology, Boston University School of Public Health, Boston University, Boston, MA, USA}
\affil[5]{Department of Global Health, Boston University School of Public Health, Boston University, Boston, MA, USA}
\affil[6]{Division of Population Medicine, School of Medicine, Cardiff University, Cardiff, UK }

\usepackage[english]{babel}
\usepackage[utf8]{inputenc}
\usepackage{csquotes}

\usepackage{chngcntr}
\usepackage{apptools}
\usepackage{amsthm}
\usepackage{multirow}

\theoremstyle{definition}

\newtheorem*{remark*}{Remark}
\newtheorem{theorem}{Theorem}[section]

\newtheorem{prop}[theorem]{Proposition}
\newtheorem{claim}[theorem]{Claim}

\newtheoremstyle{break}
  {\topsep}{0}%
  {\bfseries}{}%
  {\newline}{}%
\theoremstyle{break}
\newtheorem*{remarks*}{Remarks}
\newtheorem*{examples*}{Examples}

\makeatletter
\def\munderbar#1{\underline{\sbox\tw@{$#1$}\dp\tw@\z@\box\tw@}}
\makeatother

\usepackage{afterpage}
\usepackage{caption}

\usepackage{amsmath}
\usepackage{amsfonts}

\newcommand\independent{\protect\mathpalette{\protect\independenT}{\perp}}
\def\independenT#1#2{\mathrel{\rlap{$#1#2$}\mkern2mu{#1#2}}}

\begin{document}
\title{\huge Shall we count the living or the dead?}
\date{\today}
\maketitle
\sloppy
\thispagestyle{empty}\normalsize

\newpage

\newpage

\section*{Abstract}

In the 1958 paper ``Shall we count the living or the dead?'', Mindel C. Sheps proposed a principled solution to the familiar problem of asymmetry of the relative risk. We provide causal models to clarify the scope and limitations of Sheps' line of reasoning, and show that her preferred variant of the relative risk will be stable between patient groups under certain biologically interpretable conditions. Such stability is useful when findings from an intervention study must be generalized to support clinical decisions in patients whose risk profile differs from the participants in the study. We show that Sheps’ approach is consistent with a substantial body of psychological and philosophical research on how human reasoners carry causal information from one context to another, and that it can be implemented in practice using van der Laan et al's Switch Relative Risk, or equivalently, using Baker and Jackson's Generalized Relative Risk Reduction (GRRR). 

\section{Introduction}\label{sec1}
\subsection{Background}

When evaluating evidence in order to engage in shared decision-making about a medical intervention (such as initiating treatment with a particular drug), patients want to know the potential harms and benefits that they can expect to experience if they undergo the intervention, and if they don't undergo the intervention \citep{murray_patients_2018}. However, personalized estimates of absolute risks are rarely published, as very few studies can be powered for estimating risk in each subgroup separately \citep{cook_subgroup_2004, wang_detecting_2013}. In the absence of direct evidence for personalized risk under intervention, this can be estimated based on background information about the patient's risk profile if untreated, and a published measure of the magnitude of the effect. This procedure usually depends on a strong assumption that the relative effect of the intervention is stable across groups, i.e. on the absence of effect modification on the relative risk scale \citep{sackett2000evidence, furukawa_can_2002}. While the relative risk scale is commonly used and has attractive properties, stability of the relative risk is not a universally held belief \citep{poole_is_2015, panagiotou_commentary:_2015, spiegelman_evaluating_2017} and has unclear theoretical support. Stability of the effect measure is also a crucial consideration for randomization inference \citep{senn_controversies_2004}, and when choosing the summary parameter of a meta-analysis \citep{higgins_handbook_2011} or the link function in a general linear model.

It is well established that the relative risk is asymmetric, meaning that the predictions of a relative risk model are not invariant to recoding of the exposure variable \citep{cox_analysis_1977}. In this manuscript, we discuss a resolution to this limitation, based on an approach to modelling binary events which originated with Canadian physician and biostatistician Mindel C. Sheps (1913-1973). In \citet{sheps_shall_1958} and \citet{sheps_examination_1959}, she argued that in settings where an intervention reduces risk of an outcome, the standard risk ratio (the variant of the relative risk which is based on the probability of the outcome event, i.e.\ ``counts the dead'') is the most suitable effect measure; whereas in settings where the intervention increases risk, a variant of the relative risk which is based on the probability of \textit{not} having the event (i.e.\ ``counts the living'') is preferred. For consistency with the earlier literature, we will refer to this variant of the relative risk as the survival ratio, regardless of whether the outcome is death. While Sheps' ideas are rarely used in the applied literature, variations of her insights have been rediscovered independently multiple times \citep{ khoury_measurement_1989, bouckaert_sure_2001, furuya-kanamori_outcome_2014, huitfeldt_choice_2018, shannin_disagreement_2021}. 

This manuscript is organized as follows: In section 2, we argue that conditional stability is the centralizing consideration for choice of effect measure. In section 3, we expand upon Sheps' argument by attaching it to a causal model consistent both with standard toxicological mechanisms and with influential work on generalizability from the philosophy and psychology literature, and show that these models may lead to stability of Sheps' preferred variant of the relative risk. In section 4, we show that Sheps' recommendations can be implemented in practice by using the switch relative risk model from \citet{van_der_laan_estimation_2007}, or equivalently, using the  \citet{baker_new_2018} generalized relative risk reduction (GRRR) parameter . In section 5, we provide an example of how this line of reasoning may have implications for clinical decision making. We conclude in section 6.

\subsection{Notation}

All variables considered in the paper are defined in Table 1. We refer to the marginal risk of the outcome under the intervention as $\mu_1$ and to the corresponding ``baseline'' risk under a control condition as $\mu_0$. Because we are considering causal measures of effect, $\mu_1$ and $\mu_0$ are equated with the expectations of the potential outcomes, $Y^{a=1}$ and $Y^{a=0}$, respectively\citep{hernan_causal_2020}. When any quantity is considered in a specific subgroup, this is denoted in the subscript, e.g\ $\mu_{0, v}$. We use $\neg$ to denote the complement of an event. 

\subsection{Effect measures and effect functions}
All effect measures considered in the paper are defined in Table 2. Each effect measure defines a scale for measuring the magnitude of the contrast between $\mu_1$ and $\mu_0$. We use upper case greek letters to refer to the set of all possible values of an effect measure (and to the effect measure when considered as an unspecified parameter), and lower case greek letters to refer to a specific value of an effect measure. We use $\Lambda$ (with values $\lambda)$ whenever we need to refer to an unspecified effect measure. 

Any effect measure for a binary outcome can be represented as an effect function $g$, which takes a probability $p$ as input and outputs a real number $g_{\lambda}(p)$, often another probability. In different contexts, $p$ could be a marginal risk, e.g. $\mu_0$, or a conditional risk, e.g. $\mu_{0,l}$. Either way, the input risk relates to the control condition and the output risk relates to the corresponding risk under intervention. For example, the effect function for the odds ratio is \[ g_{\gamma} (p) = \frac{\frac{p}{1-p}\times \gamma}{1+\frac{p}{1-p}\times \gamma}\] and the (marginal) odds ratio model can be written as $\mu_1= g_\gamma (\mu_0)$. The effect function is useful because it governs many interesting features of an effect measure:

\begin{itemize}
    \item If the effect function $g_{\lambda}$ is closed on [0,1] for all $\lambda$, models based on $\Lambda$ will not produce invalid predictions
    \item If the effect function $g_{\lambda}$ is an affine transformation for all $\lambda$, $\Lambda$ is collapsible \citep{daniel_making_2021}
    \item Models based on effect measures $\Lambda$ and $\Lambda '$ are prediction-equivalent (i.e. if both models are fit in the same data, they will lead to identical predictions) if for all valid parameter values $\lambda$ in $\Lambda$, there exists a corresponding parameter value $\lambda '$ in $\Lambda '$ such that $g_{\lambda} (p)$ = $g_{\lambda '} (p)$ for all $p$
\end{itemize}

 \begin{table}
 \begin{adjustwidth}{0in}{0in}
 \caption{Definitions of variables used in the text.}
 \begin{center}

\begin{tabular}{|m{3cm}|m{3cm}|m{9cm}|} 
\hline
\textbf{Variable}& \textbf{Type} &\textbf{Remarks/Definition}\\
\hline
$A$ & Exposure & 1=Penicillin \newline 0=No Penicillin\\
\hline
$Y$ & Outcome & In examples where intervention reduces risk: 
\newline
1=Rheumatic fever \newline
0=No rheumatic fever\newline \newline In examples where intervention increases risk:
\newline 1=Anaphylaxis
\newline 0=No anaphylaxis\\
\hline
$L$ & Baseline patient characteristics & Often used as predictors of risk if untreated\\ 
\hline
$V$ & Effect modifiers & Measured determinants of group-level effect size (may be a vector of several covariates). Often a subset of $L$.\\ 
\hline
$S$ & Setting & Groups that the effect may be equal between after controlling for effect modifiers. We use lower case letters for specific values of $S$:
\newline 
$s$=Study population
\newline 
$t$=Target population\\
\hline
$B, C, D, E,Q$& Switches & Unmeasured determinants of individual response to treatment. $B,C,D$ and $E$ each have specific functions; we use $Q$ whenever we need to refer to an unspecified determinant of treatment response.\\
\hline
$\mathbb{B, C, D, E, Q}$& Switch patterns & Unmeasured combinations of switches that together determine individual treatment response.\\
\hline
$U, W$ & Other unmeasured variables &$U$ is used in models for the outcome, $W$ is used in models for the complement of the outcome.\\ 
\hline

\end{tabular}
\begin{flushleft} 
\end{flushleft} 
\end{center}
\end{adjustwidth}
 \end{table}

 \begin{table}
  \begin{adjustwidth}{0in}{0in}
 \caption{Definitions of effect measures.}
 \begin{center}
\begin{tabular}{|m{3cm}|m{5cm}|m{7cm}|} 
\hline
\textbf{} & \textbf{Effect Measure}& \textbf{Effect Function}\\
\hline
Risk difference (RD) & $\alpha = \mu_1 - \mu_0$ & $ g_{\alpha}(p)= p + \alpha $\\
\hline
Risk ratio (RR) &$\beta = \frac{\mu_1}{\mu_0}$ & $g_{\beta}(p)= p \times \beta $\\
\hline
Odds ratio (OR) & $\gamma = \frac{\sfrac{\mu_1}{(1-\mu_1)}}{\sfrac{\mu_0}{(1-\mu_0)}}$ & $g_{\gamma}(p) = \frac{(\sfrac{p\times \gamma)}{(1-p)}}{1+\sfrac{(p\times \gamma)}{(1-p)}} $ \\
\hline
Survival ratio (SR) & $\nu = \frac{1-\mu_1}{1-\mu_0}$ &$ g_{\nu}(p) = 1-(1-p)\times \nu$ \\
\hline
Switch relative risk (using 
GRRR notation) &$\theta = \begin{cases}
 1- \frac{1-\mu_1}{1-\mu_0} & \text{if } \mu_1>\mu_0 \\
 0 & \text{if } \mu_1=\mu_0 \\
 -1+\frac{\mu_1}{\mu_0} & \text{if } \mu_1<\mu_0 
 \end{cases}$ 
 & $g_{\theta}(p) = \begin{cases}
 1- (1-p) \times (1-\theta) & \text{if } \theta >0\\
 p & \text{if } \theta =0 \\
 p \times ( 1 + \theta) & \text{if } \theta <0 
 \end{cases}$\\
\hline
\end{tabular}
\begin{flushleft}
\end{flushleft} 
\end{center}
\end{adjustwidth}
 \end{table}
 
\section{Effect measure stability}\label{sec2}

Any attempt at individualizing the results from a study to a specific patient, or generalizing the results to a different population, necessarily involves invoking an explicit or implicit homogeneity assumption, possibly conditional on some set of effect modifiers. Typically, these homogeneity assumptions come in the form of a claim about conditional stability of an effect measure. 

In order to formalize how such individualization is done in practice, we will consider a situation where a clinical decision needs to be made in a patient whose risk profile can be characterized by a vector of clinically relevant risk predictors $L$. We will suppose we have access to a reasonably accurate estimate of the patient-specific baseline risk ($\mu_{0,l}$). The baseline risk is then combined with a published estimate of the magnitude of the effect ($\lambda$), in order to produce an estimate of the patient-specific risk under treatment ($\mu_{1,l}$): $\mu_{1,l}$ is equated with $g_\lambda (\mu_{0,l})$.\footnote{An example of how this procedure is used in practice can be seen in the Cochrane Handbook, which suggests that ``the risk in the intervention group (and its 95\% confidence interval) is based on the assumed risk in the comparison group and the relative effect of the intervention (and its 95\% CI)'' \citep{schunemann_chapter_2021}. In other words, the Cochrane Handbook recommends using the procedure above, and choosing the risk ratio function for $g_\lambda$).} This procedure is inherently scale-specific, and will result in different predictions for $\mu_{1,l}$ depending on choice of effect measure (see interactive figure S1). Only if $\lambda$ is stable between patient groups, will the procedure result in accurate predictions. For this reason, we give priority to conditional stability (homogeneity) above all other considerations for choice of effect measure. In appendix 1, we discuss how stability relates to other considerations for choice of effect measure 

In order to account for effect heterogeneity, the procedure may be modified to condition on a measured set of baseline covariates $V$ (where $V$ consists of those covariates that are believed to influence the magnitude of the effect, often a subset of $L$). This can be done by estimating $\lambda_{v}$ in the study, and equating $\mu_{1,l}$ with $g_{\lambda_{v}} (\mu_{0,l})$. However, this modification does not overcome the basic problem of scale-dependency, and the modified procedure will only be justified if there is reason to believe $V$ is a sufficient set of effect modifiers on the $\Lambda$ scale, i.e. that two groups from different settings that share the same value of effect modifiers $V$ will have the same effect on that scale. For example, if $\lambda$ depends only on sex, $\lambda_v$ may be expected to be equal between groups of men from different countries, and between groups of women from different countries. 

\section{Mechanisms of action} \label{sec4}

\subsection{Sufficient-component cause models}

We now proceed to outline a framework in which the procedure for choosing the conditioning set $V$ is linked to the choice of effect measure via a model for the mechanism of action, thereby facilitating a meaningful evaluation of the biological plausibility of conditional homogeneity of Sheps' preferred variant of the relative risk. These mechanisms are consistent with Patricia Cheng's power-PC framework for causal generative and preventive power \citep{cheng_covariation_1997,cartwright_natures_1989}, an approach which has considerable support in the psychology and philosophy literature \citep{hiddleston_causal_2005,glymour_causal_1998,glymour_minds_2001}, where it has been argued on both empirical and normative grounds that human reasoners use (and should use) these constructs to carry causal information from one context to another \citep{liljeholm_when_2007}. The mechanisms are also consistent with Bouckaert and Mouchart's Sure Outcomes of Random Events model \citep{bouckaert_sure_2001, mouchart_pharmacological_2019}, and with the independent joint action model from toxicology \citep{abbott_method_1925,bliss_toxicity_1939,howard_contrasting_2013}, which has previously been discussed in the epidemiological literature by Weinberg \citep{weinberg_applicability_1986, weinberg_can_2007, wacholder_inference_2011, weinberg_interaction_2012}. 

Sufficient-component cause models (``causal pie models'') \citep{rothman_causes_1976, mackie_cement_1974} are used to visualize these mechanisms. These models consider several different combinations of factors that together comprise a sufficient cause of the outcome (``causal pies''). Each pie contains component causes (``slices of the pie''), such that if every slice of any pie is present, the outcome will occur. Causal pie models can be used to show that structural knowledge about the distribution and function of the unmeasured covariates that turn treatment effects ``on'' or ``off'' can sometimes be sufficient to guarantee treatment effect stability on one specific scale, but not on others. To illustrate, we will consider the effect of Penicillin in patients with Streptococcal pharyngitis on an effectiveness outcome (rheumatic fever) and a safety outcome (anaphylaxis in the first two weeks of treatment).

While these models may not perfectly capture the underlying biology, we share the perspective of Alan Turing, who once wrote ``In this section a mathematical model  [...] will be described. This model will be a simplification and an idealization, and consequently a falsification. It is to be hoped that the features retained for discussion are those of greatest importance in the present state of knowledge''\citep{turing_chemical_1952}

\subsection{Shall we count the living or the dead?}

Like relative risks, causal pie models are not invariant to whether they represent sufficient causes of the outcome (counting the dead), or sufficient causes of not having the outcome (counting the living). Any set of conditions which together completely determine whether the outcome occurs, can be represented in either form. These model forms may differ substantially in their complexity, which we discuss further in Appendix 2.

Figure 1 shows a causal pie model for the outcome. The causal pies in this model can be partitioned into three broad classes: Class 1 contains those causal pies that do not depend on Penicillin. Class 2 contains those causal pies in which Penicillin ($A$) is a component. Class 3 contains those causal pies in which not taking Penicillin ($\neg A$) is a component. The causal pies of class 1 are taken to generate the background risk of $Y$, i.e.\ the component of the risk that occurs regardless of whether the intervention is given. We allow the distribution of the component causes of pies of class 1 to vary arbitrarily between groups, resulting in different baseline risks. We call the event that all component causes of any causal pie of class 1 is met $\mathbb{U}$. Penicillin will trigger the outcome in those people who have met every other component in at least one causal pie of class 2, and the absence of Penicillin will trigger the outcome in those who have met every other component of at least one causal pie of class 3. We will refer to the event that every non-$A$ component of at least one causal pie of class 2 is present as $\mathbb{B}$, and that every non-$\neg A$ component of at least one causal pie of class 3 is present as $\mathbb{C}$. We will refer to $\mathbb{B}$ and $\mathbb{C}$ as ``switch patterns'', the presence or absence of these switch patterns determines whether and how treatment affects the outcome in an individual. Under this model, risk under the intervention in any group $v$ will be given by $\mu_{1,v} = (\mathbb{U} \cup \mathbb{B}\vert v)$, and risk under the control condition is $\mu_{0,v} = Pr(\mathbb{U} \cup \mathbb{C}\vert v)$. 
 
Figure 2 shows a causal pie model for the complement of the outcome. Here, the causal pies can also be partitioned into three different classes: The outcome does not occur if every component cause of at least one causal pie of class 4 is met, we call this $\mathbb{W}$. If every non-intervention component cause of at least one causal pie of class 5 is met ($\mathbb{D}$), the intervention ensures that the outcome does not occur. If every non-intervention component of at least one causal pie of class 6 is met ($\mathbb{E}$), not taking Penicillin ensures that the outcome does not occur. With this causal model, risk of not having the outcome under the intervention in any group $v$ will be given by $1-\mu_{1,v} = Pr(\mathbb{W} \cup \mathbb{D}\vert v)$, and risk of not having the outcome under the control control condition is $1-\mu_{0,v} = Pr(\mathbb{W} \cup \mathbb{E}\vert v)$. 

Each switch pattern ($\mathbb{B,C,D,E}$) is associated with a characteristic effect measure (see Table 3). For any type of switch pattern, if response to treatment depends \textit{only} on one type of switch pattern, and the prevalence of the covariates that make up that switch pattern is stable between two groups, the characteristic effect measure will also be stable. In practice, treatment response usually depends on more than one type of switch pattern, but in many cases, it will be useful to identify the switch pattern type that is primarily responsible for the effect and use its characteristic effect measure as the default choice, such that effect heterogeneity can be understood as deviations from the ``pure'' mechanism that would have led to effect measure homogeneity. The central insight that we will expand upon with causal models in the remainder of this manuscript, is that we sometimes have biological reasons for believing that one type of switch pattern is predominantly responsible for the effect of the intervention, and that this has important implications both for choosing the effect measure and the conditioning set $V$. 

In addition to the modelling assumptions that are visualized in the figures, our results will depend on an assumption of monotonicity (which can either be positive, e.g.\ that Penicillin never prevents anaphylaxis in anyone who would have anaphylaxis if untreated, or negative, e.g.\ that penicillin never causes rheumatic fever in anyone who would not have it if untreated), and that the baseline risk generated by $\mathbb{U}$ or $\mathbb{W}$ is independent of the switch patterns. These are strong assumptions, which must be evaluated based on context-specific background knowledge, and accounted for in the analysis if believed to be violated.

\begin{figure}
 \caption{Sufficient causes of the outcome}
 \centering
 \begin{subfigure}{0.45\textwidth}
 \centering
\begin{tikzpicture}
\newcounter{a}
\newcounter{b}
\foreach \p/\t in {50/$U_1$, 50/$U_2$
 }
 {
 \setcounter{a}{\value{b}}
 \addtocounter{b}{\p}
 \slice{\thea/100*360}
 {\theb/100*360}
 {\p\%}{\t}
 }
\end{tikzpicture}
 \caption{Causal pie of class 1}
 \end{subfigure}
 \hfill
 \begin{subfigure}{0.45\textwidth}
 \centering
 
\begin{tikzpicture}

\foreach \p/\t in {33/$U_1$, 33/$U_3$, 34/$U_4$
 }
 {
 \setcounter{a}{\value{b}}
 \addtocounter{b}{\p}
 \slice{\thea/100*360}
 {\theb/100*360}
 {\p\%}{\t}
 }

\end{tikzpicture}

 \caption{Causal pie of class 1}
 \end{subfigure}
 \hfill
 \newline
 \begin{subfigure}{0.45\textwidth}
 \centering

\begin{tikzpicture}

\foreach \p/\t in {50/$A$, 50/$B_1$
 }
 {
 \setcounter{a}{\value{b}}
 \addtocounter{b}{\p}
 \slice{\thea/100*360}
 {\theb/100*360}
 {\p\%}{\t}
 }

\end{tikzpicture}

 \caption{Causal pie of class 2}
 
 \end{subfigure}
 \hfill
 \begin{subfigure}{0.45\textwidth}
 \centering
 
\begin{tikzpicture}

\foreach \p/\t in {33/$A$, 33/$B_2$, 34/$B_3$
 }
 {
 \setcounter{a}{\value{b}}
 \addtocounter{b}{\p}
 \slice{\thea/100*360}
 {\theb/100*360}
 {\p\%}{\t}
 }

\end{tikzpicture}

 \caption{Causal pie of class 2}
 \end{subfigure}
 \newline
 \begin{subfigure}{0.45\textwidth}
 \centering 
\begin{tikzpicture}

\foreach \p/\t in {50/$\neg A$, 50/$C_1$
 }
 {
 \setcounter{a}{\value{b}}
 \addtocounter{b}{\p}
 \slice{\thea/100*360}
 {\theb/100*360}
 {\p\%}{\t}
 }
\end{tikzpicture}
 \caption{Causal pie of class 3}
 
 \end{subfigure}
 \hfill
 \begin{subfigure}{0.45\textwidth}
 \centering
 \begin{tikzpicture}

\foreach \p/\t in {33/$\neg A$, 33/$C_2$, 34/$C_3$
 }
 {
 \setcounter{a}{\value{b}}
 \addtocounter{b}{\p}
 \slice{\thea/100*360}
 {\theb/100*360}
 {\p\%}{\t}
 }
\end{tikzpicture}
 \caption{Causal pie of class 3}
 
 \end{subfigure}
 \hfill

 \end{figure}

 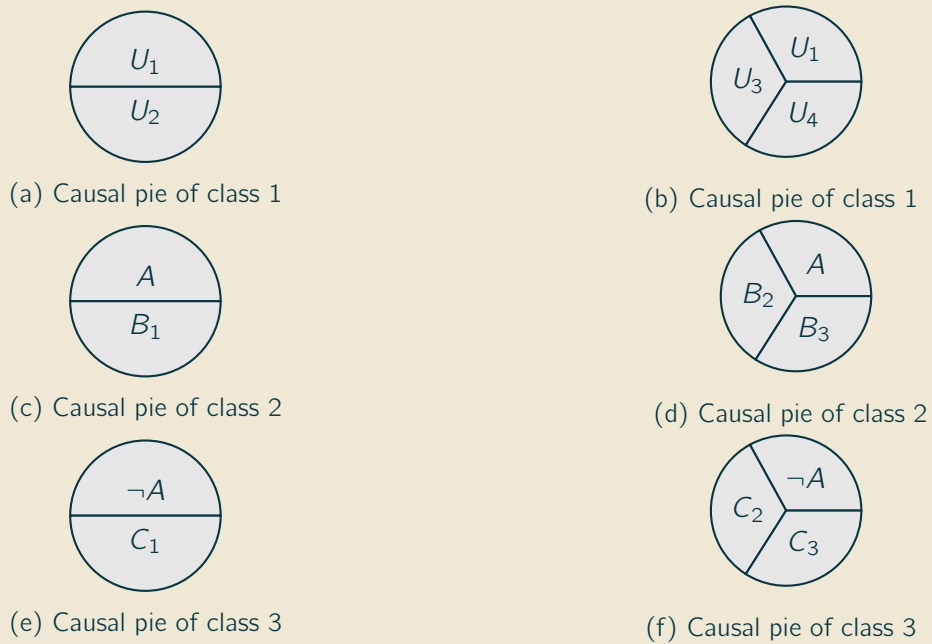
\begin{figure}
  \caption{Sufficient causes of not having the outcome}
 \centering
 \begin{subfigure}{0.45\textwidth}
 \centering
 
\begin{tikzpicture}
\foreach \p/\t in {50/$W_1$, 50/$W_2$
 }
 {
 \setcounter{a}{\value{b}}
 \addtocounter{b}{\p}
 \slice{\thea/100*360}
 {\theb/100*360}
 {\p\%}{\t}
 }
\end{tikzpicture}
 \caption{Causal pie of class 4}
 
 \end{subfigure}
 \hfill
 \begin{subfigure}{0.45\textwidth}
 \centering
 
\begin{tikzpicture}

\foreach \p/\t in {33/$W_1$, 33/$W_3$, 34/$W_4$
 }
 {
 \setcounter{a}{\value{b}}
 \addtocounter{b}{\p}
 \slice{\thea/100*360}
 {\theb/100*360}
 {\p\%}{\t}
 }

\end{tikzpicture}

 \caption{Causal pie of class 4}
 \end{subfigure}
 \newline
 \begin{subfigure}{0.45\textwidth}
 \centering
 
\begin{tikzpicture}

\foreach \p/\t in {50/$A$, 50/$D_1$
 }
 {
 \setcounter{a}{\value{b}}
 \addtocounter{b}{\p}
 \slice{\thea/100*360}
 {\theb/100*360}
 {\p\%}{\t}
 }

\end{tikzpicture}

 \caption{Causal pie of class 5}
 
 \end{subfigure}
 \hfill
 \begin{subfigure}{0.45\textwidth}
 \centering
 
\begin{tikzpicture}

\foreach \p/\t in {33/$A$, 33/$D_2$, 34/$D_3$
 }
 {
 \setcounter{a}{\value{b}}
 \addtocounter{b}{\p}
 \slice{\thea/100*360}
 {\theb/100*360}
 {\p\%}{\t}
 }

\end{tikzpicture}

 \caption{Causal pie of class 5}
 \end{subfigure}
 \newline
 \begin{subfigure}{0.45\textwidth}
 \centering 
\begin{tikzpicture}

\foreach \p/\t in {50/$\neg A$, 50/$E_1$
 }
 {
 \setcounter{a}{\value{b}}
 \addtocounter{b}{\p}
 \slice{\thea/100*360}
 {\theb/100*360}
 {\p\%}{\t}
 }
\end{tikzpicture}
 \caption{Causal pie of class 6}
 
 \end{subfigure}
 \hfill
 \begin{subfigure}{0.45\textwidth}
 \centering
 \begin{tikzpicture}

\foreach \p/\t in {33/$\neg A$, 33/$E_2$, 34/$E_3$
 }
 {
 \setcounter{a}{\value{b}}
 \addtocounter{b}{\p}
 \slice{\thea/100*360}
 {\theb/100*360}
 {\p\%}{\t}
 }
\end{tikzpicture}
 \caption{Causal pie of class 6}
 
 \end{subfigure}
 \hfill
 \end{figure}

  \begin{table}
 \begin{adjustwidth}{0in}{0in}
 \caption{Four types of switch patterns}
 \begin{center}
\begin{tabular}{ | m{4.5cm}| m{6cm}| m{3cm}| } 
\hline
\textbf{Type} & \textbf{Function of the switch pattern}& \textbf{Characteristic effect measure}\\ 
\hline
$\mathbb{B}$ (``sufficient-causal'') & $\mathbb{B}$ makes $A$ a sufficient cause of $Y$ & $SR$ \\ 
\hline
$\mathbb{C}$ (``necessary-preventive'') & $\mathbb{C}$ makes $\neg A$ a sufficient cause of $Y$ & $\frac{1}{SR}$\\ 
\hline
$\mathbb{D}$ (``sufficient-preventive'')& $\mathbb{D}$ makes $A$ a sufficient cause of $\neg Y$& $RR$\\ 
\hline
$\mathbb{E}$ (``necessary-causal'') & $\mathbb{E}$ makes $\neg A$ a sufficient cause of $\neg Y$ & $\frac{1}{RR}$\\ 
\hline
\end{tabular}

\begin{flushleft} Four types of switch patterns, and their characteristic effect measure. If treatment response is determined entirely by one class of switch patterns, the characteristic effect measure associated with that switch pattern will be stable across groups with the same prevalence of the switches.
\end{flushleft}
\end{center}
\end{adjustwidth}
 \end{table}

\subsection{Interventions that increase risk}

As an example of an intervention that increases risk, we will consider the effect of treatment with Penicillin on the risk of anaphylaxis in the first two weeks of treatment. Suppose first we relied upon the model in Figure 1. If we are additionally willing to assume positive monotonicity ($Pr(\mathbb{C})=0$) and that $\mathbb{B} \independent \mathbb{U} \vert v,s$ , it follows that the survival ratio in any group $v,s$ in the study population is equal to  $Pr(\neg \mathbb{B}\vert v, s)$:

\[ Pr(\neg \mathbb{B}\vert v,s) = Pr(\neg \mathbb{B} \vert \neg \mathbb{U}, v,s)= \frac{Pr(\neg \mathbb{B} \cap \neg \mathbb{U}\vert v,s)}{Pr(\neg \mathbb{U}\vert v,s)} = \frac{1-\mu_{1,v,s}}{1-\mu_{0,v,s}} = SR_{v,s}\]

The same logic can be used in the target population to show that under similar conditions, $Pr(\neg \mathbb{B}\vert v,t) = SR_{v,t}$. Therefore, under the model in figure 1, and with the additional assumptions stated above, $Pr(\mathbb{B} \vert v,s) = Pr(\mathbb{B} \vert t,v) \implies SR_{v,s} = SR_{v,t}$. The utility of this result arises from the observation that heterogeneity on the survival ratio scale can then be understood as resulting from biologically interpretable deviations from these conditions. In particular, heterogeneity between groups due to differences between $Pr(\mathbb{B} \vert v,s)$ and $Pr(\mathbb{B}\vert v,t)$ corresponds to the familiar concept of effect modification \citep{huitfeldt_effect_2019}, and can be accounted for by including in $V$ all predictors of the joint distribution of $B_1, B_2$ and $B_3$ whose prevalence may differ between settings $s$ and $t$. This provides a theoretical rationale for conditional effect measure stability between these settings even if the baseline risk generated by $\mathbb{U}$ differs between them. 

If instead we relied upon the model in Figure 2, a similar argument could be used to show that the reciprocal of the risk ratio (and therefore also the risk ratio) is determined by the prevalence of $\neg \mathbb{E}$, under the positive monotonicity condition $Pr(\mathbb{D})=0$ and the independence condition $\mathbb{E}\independent \mathbb{W} \vert v,s$:
 
\[ Pr(\neg \mathbb{E} \vert v,s) = Pr(\neg \mathbb{E} \vert \neg \mathbb{W}, v,s)= \frac{Pr(\neg \mathbb{E} \cap \neg \mathbb{W}\vert v,s)}{Pr(\neg \mathbb{W}\vert v,s)} = \frac{\mu_{0,v,s}}{\mu_{1,v,s}} = \frac{1}{RR_{v,s}}\]

The same argument can be made in the target population to show that $Pr(\neg \mathbb{E} \vert v,t) = RR_{v,t}$. Therefore, under this model, $Pr(\mathbb{E} \vert v,s) = Pr(\mathbb{E} \vert v,t) \implies  RR_{v,s} = RR_{v,t}$, which suggests that to account for effect heterogeneity on the risk ratio scale, it is sufficient to include in $V$ all predictors of $\mathbb{E}$ whose prevalence may differ between $s$ and $t$. Thus, a fundamental question for the choice between the survival ratio and the risk ratio for safety outcomes will be whether it will be more feasible to condition on all predictors of $\mathbb{B}$ or all predictors of $\mathbb{E}$.
 
Conditioning on all predictors of $\mathbb{B}$ may be possible in a relatively simple model of the type shown in Figure 1, in which patients exposed to Penicillin have an allergic reaction if they have gene $B_1$, or gene $B_2$ and cofactor $B_3$ which together make a person susceptible to allergic reactions. Reasoning about the predictors of these factors is sometimes a tractable task for a human reasoner. In contrast, if we instead aim to account for all predictors of $\mathbb{E}$, a much more complicated model of the type shown in Figure 2 would be required. An operative gene of type $E$ would make not taking Penicillin a sufficient cause of not getting anaphylaxis. Such a gene would essentially eliminate the possibility of having allergies to \textit{anything} if the patient can just avoid taking Penicillin, which is not realistic. While it will in theory always be possible to describe a very complicated causal pie model of this type, this would require us to incorporate the absence of every other cause of anaphylaxis as $E$-components in every pie of class 6, which will make the task of accounting for all predictors of the prevalence of $\mathbb{E}$ impossible. This leads to a preference for using the survival ratio to measure the effect. A similar argument can be used in many cases where an intervention increases risk (such as when considering adverse events), therefore, we adopt Sheps' conclusion that the survival ratio is usually a more suitable scale for interventions that increase risk. 

We note that heterogeneity may also result from violation of required conditions other than equal distribution of the switch patterns, but in many cases, it may be possible to account for this in the analysis:

\begin{itemize}
 
\item Deviations from monotonicity will occur in the model in Figure 1 if there is at least one complete set of $C$-components with prevalence greater than 0, and in the model in Figure 2 if there is at least one complete set of $D$-components with prevalence greater than 0. Monotonicity is a strong assumption, which must be evaluated separately for each exposure-outcome relationship. However, even in the absence of monotonicity, there may be approximate stability as long as the drug predominantly works in one direction (i.e.\ if there are only very few people in whom the drug works in the opposite direction from the majority). In such settings, the effect measure in all strata can be bounded using partial identification methods, and these bounds may be quite informative. \citep{cinelli_generalizing_2021}.
\item Correlation between $\mathbb{B}$ and $\mathbb{U}$ (or between $\mathbb{E}$ and $\mathbb{W}$) may occur, for example, if some people are particularly susceptible to anaphylactic reactions in general (in these people, general-factor susceptibility to allergic reactions is a component both of pies of class 1, and pies of class 2). In order to address the heterogeneity that results from this, investigators will be required to condition on markers for general-factor susceptibility to allergic reactions, or to use sensitivity analysis or partial identification methods to bound the effect.
\end{itemize}

\subsection{Interventions that reduce risk}
 
To illustrate the setting where the intervention reduces risk of outcome, we will consider the effect of Penicillin at reducing risk of rheumatic fever in patients with Streptococcal Pharyngitis. Given the model in Figure 1, the reciprocal of the survival ratio is determined by $Pr(\neg \mathbb{C}\vert v)$, under the negative monotonicity condition $Pr(\mathbb{B})=0$ and the independence condition $\mathbb{C} \independent \mathbb{U} \vert v$:

\[ Pr(\neg \mathbb{C}\vert v,s) = Pr(\neg \mathbb{C} \vert \neg \mathbb{U}, v,s)= \frac{Pr(\neg \mathbb{C} \cap \neg \mathbb{U}, v,s)}{Pr(\neg \mathbb{U}, v)} = \frac{1-\mu_{0,v,s}}{1-\mu_{1,v,s}} = \frac{1}{SR_{v,s}}\]

The same argument can be made in the target population to show that $Pr(\neg \mathbb{C} \vert v,t) = SR_{v,t}$. It therefore follows that under these conditions,  $Pr(\mathbb{C} \vert v,s) = Pr(\mathbb{C} \vert t,v) \implies  SR_{v,s} = SR_{v,t}$, and that including in $V$ predictors of $\mathbb{C}$ whose prevalence may differ between $s$ and $t$ will be sufficient to account for effect modification on the survival ratio scale.

Alternatively, using the model in Figure 2, we see that the risk ratio is determined by $Pr(\neg \mathbb{D}\vert v,s)$, under the negative monotonicity condition $Pr(\mathbb{E})=0$ and the independence condition $\mathbb{D} \independent \mathbb{W}\vert v, s$:

\[ Pr(\neg \mathbb{D}\vert v,s) = Pr(\neg \mathbb{D} \vert \neg \mathbb{W}, v,s)= \frac{Pr(\neg \mathbb{D} \cap \neg \mathbb{W}\vert v,s)}{Pr(\neg \mathbb{W}\vert v,s)} = \frac{\mu_{1,v,s}}{\mu_{0,v,s}} = RR_{v,s}\]

The same argument can be made in the target population to show that $Pr(\neg \mathbb{D}\vert v,t) = RR_{v,t}$. Therefore, under these conditions, $Pr(\mathbb{D} \vert v,s) = Pr(\mathbb{D} \vert t,v) \implies  RR_{v,s} = RR_{v,t}$, meaning that inlcuding in $V$ all predictors of $\mathbb{D}$ whose prevalence may differ between $s$ and $t$ is sufficient to account for effect modification on the risk ratio scale. 

Again, therefore, we must invoke biological knowledge to determine whether it is more feasible to condition on predictors of $\mathbb{C}$  or $\mathbb{D}$. Conditioning on all predictors of $\mathbb{D}$ may be possible in a relatively simple model of the type shown in Figure 2, in which Penicillin prevents rheumatic fever in those who a specific strain of Streptococcal Pharyngitis ($D_1$) that is highly susceptible to $\beta$-lactams, or a different strain of $\beta$-lactam susceptible Streptococcal ($D_2$) and no abnormalities of drug metabolism ($D_3$). Human reasoners can plausibly determine the potential effect modifiers which predict these. In contrast, if we rely on the model type shown in Figure 1, a much more complicated model is required. Here, factors such as $C_1, C_2$ and $C_3$ would combine to make not taking Penicillin a sufficient cause of rheumatic fever. This would mean that the patient's own immune system is irrelevant, that there could be no way to prevent the outcome other than to initiate treatment with this specific drug. Specifying a causal pie of class 3 will therefore only be possible by incorporating, as $C$-components in the model, the absence of every other potential way the body could clear the infection. This will require a very complex model containing a very large number of covariates, making it almost impossible to use predictors of $\mathbb{C}$ to reason about effect modification. For this intervention, the risk ratio model therefore appears more reasonable.

Similar logic will apply for many interventions that reduce risk. We therefore again adopt Sheps' view, that the risk ratio is preferred in the case of interventions that decrease risk (with the same caveats as in the previous section regarding monotonicity and correlation, and the same potential resolution to those caveats). However, the scope of this conclusion is more limited than the corresponding argument for interventions that increase risk, and does not apply when the outcome is all-cause mortality, as it is generally not plausible to model an intervention as being a sufficient cause of all-cause survival. We discuss this in more detail in Appendix 2. 

\subsection{Explaining biological asymmetry}

So far, we have argued that in general, switch patterns of type $\mathbb{B}$ are much more likely to explain treatment response than switch patterns of type $\mathbb{E}$, and that switch patterns of type $\mathbb{D}$ are likely to explain treatment response than switch patterns of type $\mathbb{C}$. We believe this claim matches most readers' intuition about how biological systems work. We now proceed to hint at one possible explanation for this asymmetry. For many potential interventions, our ancestors were either almost uniformly exposed or almost uniformly unexposed. For example, virtually no human ancestor was exposed to Penicillin. In such an environment, the presence of genes associated with switch pattern type $\mathbb{B}$, which causes allergy when exposed to Penicillin, will not subject the organism to any particular kind of evolutionary pressure; whereas genes associated with switch pattern type $\mathbb{E}$, which prevents all allergy in anyone who does not take Penicillin, will very quickly reach fixation (and will therefore not be plausible as a determinant of variation in treatment response). In a different but logically possible world, one in which Penicillin molecules were in the water supply, genes associated with pattern type $\mathbb{B}$ would instead have been eliminated from the gene pool, and pattern type $\mathbb{E}$ would not subject its holder to any evolutionary pressure. However, we do not live in that world, and $\mathbb{B}$ is therefore more prevalent. A similar argument could be made for the protective effect of Penicillin, by reasoning about the evolutionary pressures on bacteria, leading to a preference for models based on switch patterns of type $\mathbb{D}$ over models based on switch patterns of type $\mathbb{C}$ . 

This argument can only be applied when considering interventions for which there was a ``default'' state in the evolutionary past. It would for example not be possible to make this argument for an exposure variable such as sex, because all humans descend both from ancestors who were subjected to evolutionary pressure as men, and ancestors who were subjected to evolutionary pressure as women. Therefore, this framework does not provide a reason to expect stability of the effect of sex (and similar variables) on any scale. This again corresponds to Sheps' conclusion, that ``in this example, there is no general basis for a preference among several possible denominators''. In our view, this is not so much a shortcoming of Sheps' suggestion as a shortcoming of all effect measures: This framework provides rationale for expecting stability of the effect of some interventions but not others, the open problem of finding a stable scale for the effect of variables such as sex is left unsolved.

\subsection{An impossibility theorem for the odds ratio}

The reasoning in this section cannot be applied to the odds ratio, unless $V$ contains sufficient covariates to also guarantee equality of all effect measures. To illustrate, suppose we are able to construct a set of effect modifiers $V$ to ensure $Pr(\mathbb{Q}\vert v, s) = Pr(\mathbb{Q}\vert v, t) = Pr(\mathbb{Q} \vert v, s \cup t)$. If the odds ratio is equal between groups with the same distribution of $\mathbb{Q}$, this will then imply the following relationship:

\[
\frac{\frac{\mu_{1,v,s}}{1-\mu_{1,v,s}}}
{\frac{\mu_{0,v,s}}{1-\mu_{0,v,s}}} =
\frac{\frac{\mu_{1,v,t}}{1-\mu_{1,v,t}}}
{\frac{\mu_{0,v,t}}{1-\mu_{0,v,t}}} =
\frac{\frac{\mu_{1,v,s\cup t}}{1-\mu_{1,v,s \cup t}}}
{\frac{\mu_{0,v, s \cup t}}{1-\mu_{0,v, s \cup t}}} 
\]

This, in turn, implies that either $\mu_{1,v,s} = \mu_{1,v,t}$ and $\mu_{0,v,s} = \mu_{0,v,t}$ (i.e.\ the conditional counterfactual risks are equal between settings, which in practice means that $V$ contains every cause of the outcome) or $\mu_{1,v,s} =\mu_{0,v,s}$ and $\mu_{1,v,t} =\mu_{0,v,t}$ (i.e.\ treatment has no effect). In both cases, it follows not just that the odds ratio is stable but that \textit{every} conditional effect measure is stable between the groups $v,s$ and $v,t$. In other words, conditional stability of the odds ratio due to equal conditional distribution of individual-level determinants of treatment response can only be obtained by controlling for enough variables to also obtain conditional stability of every other effect measure. This observation is closely related to non-collapsibility \citep{daniel_making_2021}, and a similar argument can be made for any non-collapsible effect parameter.

While we caution against making overly general conclusions from this simple mathematical argument, it does demonstrate that a scientist who is selecting what effect modifiers to account for, aiming to obtain conditional homogeneity of the odds ratio, cannot be guided by biological beliefs about predictors of individual-level determinants of treatment response. We also note that \citet{doi_questionable_2020} have recently made a claim that the odds ratio is independent of baseline risks. An immediate corollary of our result is that this alleged baseline risk independence of the odds ratio cannot be a consequence of equal conditional distribution of individual-level determinants of treatment response.

\section{The switch relative risk}

Statistical modellers and clinical scientists often require an effect measure which can be specified before it is known whether the intervention increases or decreases the risk of the outcome. This motivates the switch relative risk \citep{van_der_laan_estimation_2007}, a composite effect parameter which selects a variant of the relative risk depending on whether risk of the event is higher or lower when the intervention is implemented. The switch relative risk is defined as being equal to the risk ratio if the intervention reduces risk of the outcome, and equal to survival ratio if the intervention increases risk of the outcome.

\citet{baker_new_2018} proposed a notationally convenient representation of the switch relative risk, which they referred to as the ``generalized relative risk reduction (GRRR)'' and gave the symbol $\theta$. GRRR is prediction-equivalent to the switch relative risk in the sense defined in section 1.3, and is defined as being equal to one minus the survival ratio if the intervention increases risk of the outcome, equal to 0 if the intervention has no effect, and equal to the risk ratio minus one if the intervention reduces risk: 

\[\theta = \begin{cases}
 1- \frac{1-\mu_1}{1-\mu_0} & \text{ if } \mu_1>\mu_0 \\
 0 & \text{ if } \mu_1=\mu_0 \\
 -1+\frac{\mu_1}{\mu_0} & \text{ if } \mu_1<\mu_0 
 \end{cases}\]

The effect function of $\theta$ is its inverse with respect to $\mu_1$:

\[
g_{\theta} (p) = \begin{cases}
 1- (1-p) \times (1-\theta) & \text{ if } \theta >0\\
 p & \text{ if } \theta =0 \\
 p \times ( 1 + \theta) & \text{ if } \theta <0 
 \end{cases}
\]

To illustrate calculation of $\theta$ from data, suppose an RCT shows that risk in the control group is $2\%$ and risk in the intervention group is $1\%$. Then, $\theta = -1+\frac{0.01}{0.02} = -0.5$. If instead risk in the control group is $2\%$ and risk in the intervention group is $4\%$, $\theta = 1-\frac{0.96}{0.98} \approx 0.02$. In general, the causal $\theta$-parameter will be in the range in $[-1,1]$, and will be positive if treatment increases risk, negative if treatment reduces risk, closer to 0 if effects are small and closer to $1$ or $-1$ if effects are large. 

If we have information on the baseline risk in the group that our patient belongs to, and wish to combine this with a published estimate of $\theta$ in order to predict their risk of the outcome under the intervention, this can be calculated using the effect function. To illustrate, if a doctor believes that her patient belongs to a group whose baseline risk is $3\%$, and is told that $\theta = 0.02$, she will predict that risk under the intervention will be $1-(1-0.03)\times (1-0.02) \approx 5\%$. If she is instead told that $\theta=-0.5$, she will predict that the patient's risk under the intervention is $0.03\times 0.5=1.5\%$. The effect function is closed on the interval $[0,1]$, this procedure will therefore not result in predicting invalid probabilities.

 Fig. 3 illustrates the $\theta$ scale on a number line. In somewhat of an oversimplification, if we assume that the intervention only works in one direction (monotonicity), a positive causal $\theta$ can be interpreted as the probability of ``outcome changing'' in response to treatment among those who would not experience the outcome if untreated, and the absolute value of a negative causal $\theta$ can be interpreted as the probability of outcome changing in response to treatment among those who would have experienced the outcome if untreated \citep{huitfeldt_choice_2018}. These probabilities are closely related to sufficiency scores, which differ only slightly in the counterfactual definition of the conditioning event, and which have recently been argued to improve upon state-of-the-art approaches to explainability of artificial intelligence \citep{galhotra_explaining_2021}. 
 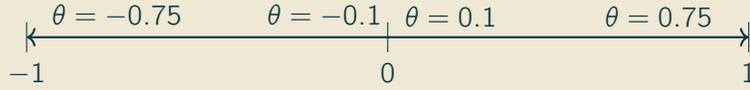
\begin{figure}
\begin{center}
\begin{tikzpicture}[xscale = 1.2]

\draw [thick, <->] (-4,0) -- (4,0);

\draw (0,-0.2) -- (0,0.2); 
\draw (4,-0.2) -- (4,0.2);
\draw (-4,-0.2) -- (-4,0.2);
\put(-90,0){\circle{4}}
\put(90.5,0){\circle{4}}
\put(10,0){\circle{4}}
\put(-10,0){\circle{4}}

\node [below] at (0,-0.2) {$0$};
\node [below] at (-4,-0.2) {$-1$};
\node [below] at (4,-0.2) {$1$};
\node[coordinate,label=above:{$\theta=-0.75$}] at (-3,0) {};
\node[coordinate,label=above:{$\theta=-0.1$}] at (-0.7,0) {};
\node[coordinate,label=above:{$\theta=0.1$}] at (0.7,0) {};
\node[coordinate,label=above:{$\theta=0.75$}] at (3,0) {};
\end{tikzpicture}
\caption{Number line displaying the $\theta$ scale. $\theta=-0.75$ is a large protective effect corresponding to a risk ratio of 0.25. $\theta=-0.1$ is a moderate protective effect corresponding to a risk ratio of 0.9. $\theta=0.1$ is a moderate harmful effect corresponding to a survival ratio of 0.9. $\theta=0.75$ is a large harmful effect corresponding to a survival ratio of 0.25}
\end{center}
\end{figure}

We note that the switch relative risk is a disjunctive effect measure, therefore, it may be challenging to give it a realist interpretation. A realist interpretation of the effect measure itself is however not necessary; as we have shown, stability of the switch relative risk is simply a useful mathematical consequence of certain underlying biological structures (which may, or may not, be given a realist interpretation). 

\subsection{Collapsibility}
The effect function for the switch relative risk is an affine transformation for any parameter value. The parameter is therefore collapsible, and there exist weights such that the marginal causal effect measure is equal to a weighted average of the conditional causal effect measures \citet{huitfeldt_collapsibility_2019}. In this setting, we note that the conditional switch relative risk $\theta_v$ is defined as:
\[\theta_v = \begin{cases}
 1- \frac{1-\mu_{1,v}}{1-\mu_{0,v}} & \text{ if } \mu_{1,v}>\mu_{0,v} \\
 0 & \text{ if } \mu_{1,v}=\mu_{0,v} \\
 -1+\frac{\mu_{1,v}}{\mu_{0,v}} & \text{ if } \mu_{1,v}<\mu_{0,v} 
 \end{cases}\]

We next proceed to provide $\left\{w_v:v\in{V}\right\}$, with each $w_v\geq0$ and $\sum_{v\in{V}}w_v=1$ such that $$\theta=\sum_{v\in{V}}w_v\theta_v.$$ 

Proposition: The collapsibility weights for $\theta$ are $$w_v=\frac{\theta^+-\sum_{v\in{V}}w^*_v\theta_v}{\theta^+-\theta^-}w^-_v+\frac{\sum_{v\in{V}}w^*_v\theta_v-\theta^-}{\theta^+-\theta^-}w^+_v$$  where 
$$\delta=\sum_{v:\theta_v<0}\mu_{0,v}p_v\theta_v+\sum_{v:\theta_v\geq0}(1-\mu_{0,v})p_v\theta_v$$
$$w^*_v=\frac{\sum_{v:\theta_v<0}\mu_{0,v}p_v+\sum_{v:\theta_v\geq0}(1-\mu_{0,v})p_v}{I(\delta<0)\mu_0+I(\delta\geq0)(1-\mu_0)}$$
$$\mu_0^-=\sum_{v:\theta_v<0}\mu_{0,v}p_v,$$
$$\mu_0^+=1-\sum_{v:\theta_v\geq0}(1-\mu_{0,v})p_v,$$
$$w_v^-=\frac{I(\theta_v<0)\mu_{0,v}p_v}{\mu_0^-},$$
$$w_v^+=\frac{I(\theta_v\geq0)(1-\mu_{0,v})p_v}{1-\mu_0^+}.$$
$$\theta^-=\sum_{v\in{V}}w_v^-\theta_v$$ and $$\theta^+=\sum_{v\in{V}}w_v^+\theta_v$$

Proof: See appendix 4

\subsection{Relative benefits and absolute harms}

If the switch relative risk is found to be too cumbersome for practical use, Sheps' recommendations can often be approximated with a careful choice from standard effect measures. When the outcome is rare, the risk difference is closely approximated by one minus the survival ratio, and is therefore also nearly stable. This justifies individualizing treatment based on ``relative benefits and absolute harms'', as previously suggested by leading practitioners of evidence-based medicine \citep{glasziou_evidence_1995,lubsen_large_1989}.

\section{A practical example}\label{sec5}

Sheps' recommendations are generally equivalent to the standard approach when considering the primary effectiveness outcome of an intervention, but would result in a clinically meaningful change in how empirical evidence is used to inform predictions about the risk of adverse events. This can have substantial implications in settings where a clinician must determine whether the predicted benefits outweigh the predicted harms for patients whose risk profile differs from the typical participant in the study. To illustrate, we will consider the Pfizer BNT162b2 mRNA Covid-19 Vaccine, which has been shown to have an effectiveness of 95 percent \citep{polack_safety_2020} at preventing the original strain of Covid-19, corresponding to $\theta=0.95$ or $RR=0.05$. The effectiveness of the vaccine was presented on a scale that is consistent with Sheps' recommendations, following her advice would therefore not alter any predictions about the benefits of vaccination.

For questions about safety, this will not be the case. For example, a nationwide study in Israel has shown that the vaccine is associated with a small but possibly relevant elevated risk of Myocarditis. \citet{barda_safety_2021} reported this in terms of a risk ratio of 3.2 . Taking this result at face value, a clinician with a patient who has a baseline risk of Myocarditis of 1\% (significantly higher than the population average, perhaps reflecting a history of HIV infection or other prognostic factors for myocarditis) would conclude that the patient has a 3.2\% risk of myocarditis if given the vaccine. Depending on the risk of infection if unvaccinated, and depending on the availability of other vaccines, this may well lead to a determination that the harms of vaccination outweigh the benefits for this particular patient. 

If the results from Barda et al had instead been presented in terms of the switch relative risk or the survival ratio ($\theta=0.000027$, SR=0.999973), as Sheps would have recommended, this would enable the clinician to conclude that the risk of myocarditis changes from 1\% to approximately 1.0027\% when the patient is vaccinated. We would argue that this approach leads to a much more realistic estimate, consistent with a biologically interpretable hypothesis that approximately 0.0027\% percent of the population carry some form of a ``switch'' that makes them susceptible to myocarditis if vaccinated. This hypothesis may not perfectly describe the underlying biology, it is for example possible that the presence of this switch is correlated with baseline risk of myocarditis, in which case a sensitivity analysis is needed to explore the potential consequences of such correlation. We maintain that even the upper bounds of this sensitivity analysis is unlikely to produce risk estimates as high as what one would obtain if the analysis relied on homogeneity of the risk ratio or the odds ratio. In our view, Sheps' approach provides a starting point for reasoning about what complications must be accounted for in the analysis, in order to meaningfully summarize the risk of adverse effect on a numerical scale.

\section{Discussion}\label{sec6}

\subsection{Limitations}

We have presented a purely theoretical argument for approximate stability of a specific variant of the relative risk, in some situations where the joint distribution of unmeasured determinants of treatment response is reasonably expected to be approximately stable. Ideally, our argument would be supported by empirical evidence. Empirical evaluation of relative stability of different measures of effect is not theoretically straightforward, as standard tests for homogeneity have different power for different measures of effect \citep{poole_is_2015}. We note that earlier literature contains convincing empirical evidence for stability of the risk ratio in settings where the intervention reduces risk \citep{deeks_issues_2002}. Testing the empirical stability of the survival ratio for interventions that increase risk of the outcome should be a priority for future work.

It is not always possible to convincingly establish stability of any effect measure. If this is not possible, it may instead be necessary to aim for conditional stability of counterfactual distributions across populations, in order to allow generalizability of $\mu_{1,v}$ rather than $\lambda_v$. This is a much more ambitious undertaking, and will require investigators to account for all causes of the outcome whose distribution may differ across populations \citep{vanderweele_confounding_2012,pearl_external_2014}.

\subsection{Conclusions}

Under certain biologically interpretable assumptions about the distribution and function of switch patterns that turn the effect of treatment ``on'' or ``off'', the conditional survival ratio will be stable between different settings if the intervention increases risk of the outcome, and the conditional risk ratio will be stable between settings if the intervention reduces risk of the outcome. This supports the recommendations from Sheps' landmark 1958 paper ``Shall we count the living or the dead?'' and motivates the switch relative risk, which becomes the survival ratio if the intervention increases risk of the outcome, and becomes the risk ratio if the intervention reduces risk of the outcome. 

The models which justify these conclusions are consistent with Cheng's theory of generative and preventive causal power and with the independent action model from toxicology; the conditions which lead to stability of Sheps' preferred variant of the relative risk are thus better understood than for any other measure of effect. While the model will rarely be a perfect description of reality, an advantage of linking the choice of effect measure to a causal mechanism is that effect measure heterogeneity can then be understood as biologically interpretable deviation from the mechanism, which may lead to clearer reasoning about how to account for potential effect measure modification when generalizing experimental findings to patients whose risk profile differs from the participants in the study.

\bigskip
\begin{center}
{\large\bf SUPPLEMENTARY MATERIAL}
\end{center}

\section*{Acknowledgements}
The authors thank Marco Piccininni, Lyle Gurrin, Stephen Senn, Carlos Cinelli, Tim Morris, Mats J Stensrud and Daniel Farewell for helpful comments on earlier drafts of this manuscript. Dr. Daniel acknowledges support from a Sir Henry Dale Fellowship jointly funded by the Wellcome Trust and the Royal Society (grant number 107617/Z/15/Z) 

\section*{Appendix 1: Other considerations for choice of effect measure}

\subsection*{Decision theory}

Rational clinical decision making depends not only on the decision maker's beliefs about the risk of the outcome under treatment and under the control condition, but also on the patient's utility function (or the collective social welfare function), i.e.\ a monotonically increasing function over whatever outcome the decision maker is optimizing for, reflecting their values and risk posture \citep{kreps_notes_1988}. Information about the utility function can usually be elicited by asking patients (for example using standard gambles \citep{farquhar_utility_1984}). 

If the utility function is risk neutral with respect to the number of survivors (i.e.\ if the second derivative of the utility function over number of survivors is zero), the same decision will always be made in two groups between whom the baseline risks differ but the risk difference is equal. To illustrate, if we know that an intervention reduces risk by one percentage point in all groups, and the baseline risk in group $v$ is $1\%$ and the baseline risk in group $v'$ is $99\%$, then these two groups will get the ``same benefit'' in terms of how many additional people will survive if given the intervention. For this reason, if a clinician needs to make a decision about whether the benefits outweigh the harms of an intervention in a patient with neutral risk posture, the risk difference is \textit{in theory} the only input to the decision problem that needs to be known. 

However, we believe it is rare that a decision maker will find themselves in a situation where this property can be utilized. A risk difference that is applicable to the patient will only be known if both $\mu_{0,l}$ and $\mu_{1,l}$ are known (in which case the risk difference is superfluous, as the decision maker can use the absolute risks, which contain strictly more information and allow arbitrary utility functions) or if the risk difference itself is known to be stable (which would be convenient, but such convenience is very poor justification for choosing to rely upon it for generalizability). 

\subsection*{Interpretability}

It has been argued that interpretability is a key consideration for choice of effect measure, in part because this is central to explainability of the algorithms used for decision making. We recognize that interpretability is a desirable feature of an effect measure, but we note that decision makers have little use for an intuitive interpretation of the effect measure in the study population if the same interpretation does not also validly explain how the intervention will affect the patient. Therefore, in the absence of any argument for stability, interpretability can not be a primary consideration. 

\section*{Appendix 2: Causal pie models}

\subsection*{Models for survival}
Some readers may be troubled by doubts about whether the sufficient-component cause model can be applied to the absence of the outcome event. For example, the textbook Modern Epidemiology \citep{rothman_modern_2008} argues that ``Sheps (1958) once asked, ``Shall we count the living or the dead?''. Death is an event, but survival is not. Hence, to use the sufficient-component cause model, we must count the dead. This model restriction can have substantive implications''. 

We do not accept this restriction. In theory, any complete list of causal pies leading to the event can be restated as a complete list of causal pies that are sufficient causes of not having the event. These two models are therefore different representations of the same underlying process, each model will be fully valid if used appropriately, and each may be useful for making a different methodological point. However, it will usually not be realistic to represent the intervention as a component in a sufficient cause of survival: this would mean that the intervention would prevent even unrelated causes of death. We believe this observation accounts for at least some intuitive discomfort with considering survival as the outcome event.

This does not necessarily mean that such models are unrealistic for more restricted non-event outcomes, such as non-incidence of a specific disease. It is often entirely plausible to represent an intervention as a component of a sufficient cause of not experiencing a more restricted outcome, such as rheumatic fever. For these reasons, our examples intentionally relate to settings where the outcome of interest is short-term incidence of a specific disease rather than all-cause mortality. In many cases where this model is plausible, for example when the intervention is sufficient cause of an absorbing state in which the patient is no longer at risk from the outcome, human language allows us to talk about the negative outcome event (in which the patient survived) as a positive event, for example the patient was ``cured'' or ``recovered from disease''.

\subsection*{Model choice and invalid predictions}

We note that if a model states that $E$ makes $\neg A$ a sufficient cause of $\neg Y$, and the data implies that the baseline risk that is higher than the prevalence of $\neg E$, then the model is inconsistent with observations. This phenomenon is closely related to the fact that multiplicative models sometimes result in predictions outside the range of valid probabilities. In general, models that are based on switches of type $B$ and $D$ are consistent with any baseline risk but may be falsified by some values of risk under treatment; whereas models that are based on $C$ and $E$ are consistent with any risk under treatment but may be falsified by some values of baseline risk.

\section*{Appendix 3: Graphical Models}

In this appendix, we show that an identical argument for stability of Sheps' preferred variant of the relative risk can be made using directed acyclic graphs rather than causal pie models.

 The reasoning outlined in the main manuscript depends upon counterfactual independence relations of the type $Y^{a=1} \amalg S \vert Y^{a=0}, V$. On traditional causal graphs, such independence relations cannot immediately be inferred, as the graphs do not contain separate nodes for the counterfactuals $Y^{a=1}$ and $Y^{a=0}$. Recently, Cinelli and Pearl \citep{cinelli_generalizing_2021} introduced a graphical approach that enables reasoning about such independence conditions, by showing the counterfactuals on the graph. To illustrate a simplified version of this idea, we will first consider a simple example of how such a graph might look. A naive attempt to draw a causal graph with nodes for counterfactuals is shown in Figure 4:

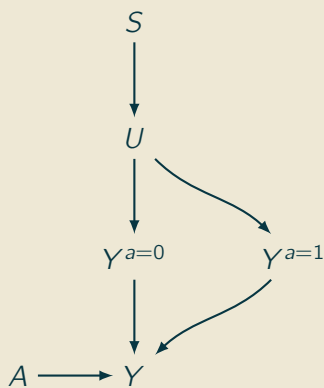
\begin{figure}
\begin{tikzpicture}
 \node (1) {$S$};
 \node [below =of 1] (2) {$U$};
 \node [below =of 2] (5) {$Y^{a=0}$};
 \node [right =of 5] (6) {$Y^{a=1}$};
 \node [below =of 5] (7) {$Y$};
\node [left =of 7](8) {$A$};
 \draw[Arrow, thick] (1) to [out=270, in=90] (2);
   \draw[Arrow, thick] (2) to [out=315, in=135] (6);
\draw[Arrow, thick] (6) to [out=225, in=45] (7);
\draw[Arrow, thick] (5) to [out=270, in=90] (7);
\draw[Arrow, thick] (2) to [out=270, in=90] (5);
\draw[Arrow, thick] (8) to [out=0, in=180] (7);
\end{tikzpicture}
\caption{Causal DAG displaying counterfactuals $Y^{a=0}$ and $Y^{a=1}$ separately}
\end{figure}

Consider the nodes representing anaphylaxis under the intervention ($Y^{a=1}$) and anaphylaxis under the control condition ($Y^{a=0}$). In most settings, $Y^{a=1}$ and $Y^{a=0}$ share most of their causes. The true graph therefore almost certainly has dense connections between these two nodes, running via the node labelled $U$. Trying to measure sufficient covariates in order to block all paths that result in D-connection between $Y^{a=1}$, $Y^{a=0}$ and $S$ would normally be hopeless. But let us next consider settings where we are additionally willing to impose constraints that arise from our background knowledge that $Y^{a=1}$ and $Y^{a=0}$ are very closely related constructs, for example such that $Y^{a=1}$ is set to $Y^{a=0}$ unless a specific covariate is present. This covariate then acts like a switch, it turns on or off the effect of $A$. If only one such type of switch is present, the graph in Figure 4 can be simplified. For example, if the effect of $A$ on $Y$ depends only of switches of type $B$, the entire assignment mechanism for $Y^{a=1}$ can be specified with a graph where its only parents are $Y^{a=0}$ and the switch (see Figure 5). On such a graph, if we condition on sufficient variables $V$ to block all paths between the switch and the population indicator $S$, we can read off the independence condition $Y^{a=1} \amalg S \vert Y^{a=0}, V$, which will play a key role in analysis of stability of effect measures, since in combination with a monotonicity assumption, it ensures that if the intervention increases risk, conditioning on $V$ is sufficient for the survival ratio to be stable across populations $S$. In the absence of monotonicity, there may be approximate stability, within bounds that may be quite informative. 

Now consider a possible world where instead of there only being switches of type $B$, there were only switches of type $E$. In such a world, the data generating mechanism would be described by the graph in Figure 6. Now, the analysis is reversed, and the risk ratio will be stable across groups. This raises an obvious question: Why would an investigator assume that the true data generating mechanism is better described by Figure 5 than by Figure 6? We argue that the answer to this lies in the same background knowledge as we discussed in the previous sections: Switches of type type $B$ are often more plausible than switches of type $E$, leading to a preference for Figure 5. 

This analysis can be generalized to some settings where there are multiple types of switches. For example, if the effect of $A$ on $Y$ depends only on switches of $B$ and $D$, the assignment mechanism for $Y^{a=1}$ can be represented as depending only on the node for $Y^{a=0}$ and on $B$ and $D$. However, if the assignment mechanism depends on both switches of type $B$ and switches of type $E$, our specification leads to paradoxical circular assignment, with a bidirectional arrow between $Y^{a=1}$ and $Y^{a=0}$, complicating any attempt to infer independences of relevance to effect measure stability. In general, switches of type $B$ are coherent with switches of type $D$, and switches of type $C$ are coherent with switches of type $E$. If background knowledge suggests that two incoherent types of switches play a significant role, these approaches will not be applicable.

\pagebreak
\begin{figure}
\begin{tikzpicture}
 \node (1) {$V$};
 \node [right=of 1] (2) {$S$};
 \node [below =of 1] (3) {$B$};
 \node [below =of 2] (4) {$U$};
 \node [below =of 3] (5) {$Y^{a=1}$};
 \node [below =of 4] (6) {$Y^{a=0}$};
 \node [below =of 5] (7) {$Y$};
\node [left =of 7](8) {$A$};

 \draw[Arrow, thick] (1) to [out=270, in=90] (3);
  \draw[Arrow, thick] (2) to [out=180, in=0] (1);
 \draw[Arrow, thick] (2) to [out=270, in=90] (4);

  \draw[Arrow, thick] (3) to [out=270, in=90] (5);
   \draw[Arrow, thick] (5) to [out=270, in=90] (7);

\draw[Arrow, thick] (6) to [out=225, in=45] (7);
\draw[Arrow, thick] (6) to [out=180, in=0] (5);
\draw[Arrow, thick] (4) to [out=270, in=90] (6);
\draw[Arrow, thick] (8) to [out=0, in=180] (7);
\end{tikzpicture}
\caption{Causal DAG where effect of $A$ on $Y$ depends only on switches of type $B$}
\end{figure}
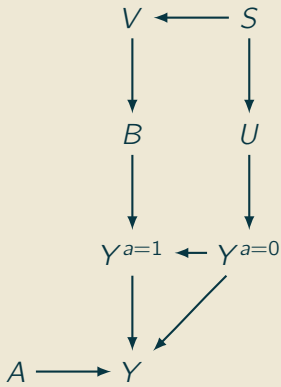
\begin{figure}
\begin{tikzpicture}
 \node (1) {$S$};
 \node [right=of 1] (2) {$V$};
 \node [below =of 1] (3) {$W$};
 \node [below =of 2] (4) {$E$};
 \node [below =of 3] (5) {$Y^{a=1}$};
 \node [below =of 4] (6) {$Y^{a=0}$};
 \node [below =of 5] (7) {$Y$};
\node [left =of 7](8) {$A$};

 \draw[Arrow, thick] (1) to [out=270, in=90] (3);
  \draw[Arrow, thick] (1) to [out=0, in=180] (2);
 \draw[Arrow, thick] (2) to [out=270, in=90] (4);

  \draw[Arrow, thick] (3) to [out=270, in=90] (5);
   \draw[Arrow, thick] (5) to [out=270, in=90] (7);

\draw[Arrow, thick] (6) to [out=225, in=45] (7);
\draw[Arrow, thick] (5) to [out=0, in=180] (6);
\draw[Arrow, thick] (4) to [out=270, in=90] (6);
\draw[Arrow, thick] (8) to [out=0, in=180] (7);
\end{tikzpicture}
\caption{Causal DAG where effect of $A$ on $Y$ depends only on switches of type $E$}
\end{figure}
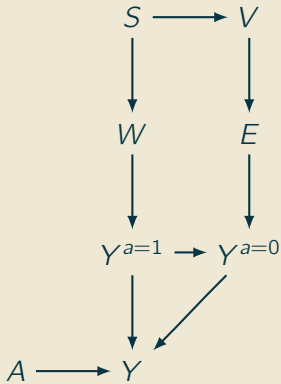

\section*{Appendix 4: Collapsibility weights for the switch risk ratio}\label{app:collapsibility_weights}
Suppose $A$ is a binary exposure, $Y$ a binary outcome and $V$ a discrete covariate, not affected by $A$. Write $\textnormal{Pr}(V=v)=p_v, \; v\in{V}$. Write the conditional and marginal potential outcome risks as follows:
\begin{align*}
    \mu_{a,v}&=\textnormal{Pr}\left\{Y^{a}=1|V=v\right\}\\    \mu_a&=\textnormal{Pr}\left\{Y^{a}=1\right\}
\end{align*}
We can then write the conditional switch relative risk parameters, $\left\{\theta_v:v\in{V}\right\}$, as the parameters that satisfy
$$\mu_{1,v}=\left\{\begin{array}{ll}\mu_{0,v}(1+\theta_v)&\textnormal{\ if\ }\theta_v<0\\1-(1-\mu_{0,v})(1-\theta_v)&\textnormal{\ if\ }\theta_v\geq 0\end{array}\right.$$
and the marginal switch relative risk parameter $\theta$ as the parameter that satisfies: 
$$\mu_1=\left\{\begin{array}{ll}\mu_0(1+\theta)&\textnormal{\ if\ }\theta<0\\1-(1-\mu_0)(1-\theta)&\textnormal{\ if\ }\theta\geq 0\end{array}\right.$$

\begin{prop} The switch relative risk is a collapsible effect measure in the sense defined by Huitfeldt \emph{et al.}\ (2019). That is, there exist weights $\left\{w_v:v\in{V}\right\}$, with each $w_v\geq0$ and $\sum_{v\in{V}}w_v=1$ such that $$\theta=\sum_{v\in{V}}w_v\theta_v.$$ 
\end{prop}

\begin{proof}
Let $$\delta=\sum_{v:\theta_v<0}\mu_{0,v}p_v\theta_v+\sum_{v:\theta_v\geq0}(1-\mu_{0,v})p_v\theta_v$$
where throughout $\sum_{v\in\emptyset}\cdot=0$, and let
$$w^*_v=\frac{\sum_{v:\theta_v<0}\mu_{0,v}p_v+\sum_{v:\theta_v\geq0}(1-\mu_{0,v})p_v}{I(\delta<0)\mu_0+I(\delta\geq0)(1-\mu_0)}$$
where $\mu_0=\sum_{v\in{V}}\mu_{0,v}p_v$. 

Then $$\sum_{v\in{V}}w^*_v\theta_v=\theta$$ by Claim \ref{claim2} below, but it does not follow necessarily that $\sum_{v\in{V}}w^*_v=1$. In particular, whenever there is at least one negative and at least one positive $\theta_v$, then $\sum_{v\in{V}}w^*_v\neq 1$.

Thus we define the following: $$\mu_0^-=\sum_{v:\theta_v<0}\mu_{0,v}p_v,$$
$$\mu_0^+=1-\sum_{v:\theta_v\geq0}(1-\mu_{0,v})p_v,$$
$$w_v^-=\frac{I(\theta_v<0)\mu_{0,v}p_v}{\mu_0^-},$$
and
$$w_v^+=\frac{I(\theta_v\geq0)(1-\mu_{0,v})p_v}{1-\mu_0^+}.$$

Note that $w_v^-$ and $w_v^+$ are undefined above if all $\theta_v$ are positive or all negative, respectively, and thus we add that
$$w_v^-=0\;\forall v\textnormal{\ if\ }\theta_v\geq 0\;\forall v$$
and
$$w_v^+=0\;\forall v\textnormal{\ if\ }\theta_v<0\;\forall v.$$

Next we define $$\theta^-=\sum_{v\in{V}}w_v^-\theta_v$$ and $$\theta^+=\sum_{v\in{V}}w_v^+\theta_v$$
and finally we let
$$w_v=\frac{\theta^+-\theta}{\theta^+-\theta^-}w^-_v+\frac{\theta-\theta^-}{\theta^+-\theta^-}w^+_v$$
where $\theta=\sum_{v\in{V}}w^*_v\theta_v$.

By Claims \ref{claim3}--\ref{claim5} below, we have that $\sum_{v\in{V}}w_v\theta_v=\theta$, that $\sum_{v\in{V}}w_v=1$ and that $w_v\geq 0 \;\forall v$, which completes the proof.
\end{proof}

\begin{claim}\label{claim2} $\sum_{v\in{V}}w^*_v\theta_v=\theta$.
\end{claim}
\begin{proof}
First note that $\delta<0 \Leftrightarrow \sum_{v\in{V}}w^*_v\theta_v<0$. Next, if $\delta<0$ then
\begin{align*}
\mu_0\left(1+\sum_{v\in{V}}w^*_v\theta_v\right)&=
\mu_0+\sum_{v:\theta_v<0}\mu_{0,v}p_v\theta_v+\sum_{v:\theta_v\geq0}(1-\mu_{0,v})p_v\theta_v\\
&=
\sum_{v:\theta_v<0}\left(\mu_{0,v}p_v+\mu_{0,v}p_v\theta_v\right)+\sum_{v:\theta_v\geq0}\left\{\mu_{0,v}p_v+(1-\mu_{0,v})p_v\theta_v\right\}\\
&=
\sum_{v:\theta_v<0}\mu_{0,v}\left(1+\theta_v\right)p_v+\sum_{v:\theta_v\geq0}\left\{1-\left(1-\mu_{0,v}\right)\left(1-\theta_v\right)\right\}p_v\\
&=
\sum_{v\in{V}}\mu_{1,v}p_v\\
&=\mu_1
\end{align*}
and if $\delta\geq 0$ then
\begin{align*}
1-\left(1-\mu_0\right)\left(1-\sum_{v\in{V}}w^*_v\theta_v\right)&=
\mu_0+\sum_{v:\theta_v<0}\mu_{0,v}p_v\theta_v+\sum_{v:\theta_v\geq0}(1-\mu_{0,v})p_v\theta_v\\
&=\mu_1.
\end{align*}
Thus $\sum_{v\in{V}}w^*_v\theta_v=\theta$.
\end{proof}

\begin{claim}\label{claim3} $\sum_{v\in{V}}w_v\theta_v=\theta$.
\end{claim}
\begin{proof}
\begin{align*}\sum_{v\in{V}}w_v\theta_v&=
\frac{\theta^+-\theta}{\theta^+-\theta^-}\sum_{v\in{V}}w^-_v\theta_v+\frac{\theta-\theta^-}{\theta^+-\theta^-}\sum_{v\in{V}}w^+_v\theta_v
\\&=
\frac{\theta^+-\theta}{\theta^+-\theta^-}\theta^-+\frac{\theta-\theta^-}{\theta^+-\theta^-}\theta^+
\\&=\theta\end{align*}
\end{proof}

\begin{claim}\label{claim4} $\sum_{v\in{V}}w_v=1$.
\end{claim}
\begin{proof}
\begin{align*}\sum_{v\in{V}}w_v&=
\frac{\theta^+-\theta}{\theta^+-\theta^-}\sum_{v\in{V}}w^-_v+\frac{\theta-\theta^-}{\theta^+-\theta^-}\sum_{v\in{V}}w^+_v
\\&=\frac{\theta^+-\theta}{\theta^+-\theta^-}+\frac{\theta-\theta^-}{\theta^+-\theta^-}
\\&=1\end{align*}
since $\sum_{v\in{V}}w^-_v=\sum_{v\in{V}}w^+_v=1$.
\end{proof}

\begin{claim}\label{claim5} $w_v\geq 0 \;\forall v$.
\end{claim}
\begin{proof}
It is clear that $w^-_v\geq 0$ and $w^+_v\geq 0 \;\forall v$, and so it remains to be shown that $\theta^-\leq\theta\leq\theta^+$.

Since $\theta^-\leq 0$ and $\theta^+\geq 0$, it suffices to show that, if $\theta<0$ then $\theta\geq \theta^-$ and if  $\theta\geq0$ then $\theta\leq \theta^+$.

First suppose $\theta<0$. Then 
\begin{align*}
\theta-\theta^-&=\frac{\sum_{v:\theta_v<0}\mu_{0,v}p_v\theta_v+\sum_{v:\theta_v\geq0}(1-\mu_{0,v})p_v\theta_v}{\mu_0}-\frac{\sum_{v:\theta_v<0}\mu_{0,v}p_v\theta_v}{\sum_{v:\theta_v<0}\mu_{0,v}p_v}\\
\Rightarrow\;\mu_0\left(\theta-\theta^-\right)&=\left(1-\frac{\mu_0}{\sum_{v:\theta_v<0}\mu_{0,v}p_v}\right)\sum_{v:\theta_v<0}\mu_{0,v}p_v\theta_v+\sum_{v:\theta_v\geq0}(1-\mu_{0,v})p_v\theta_v
\\&=\frac{\sum_{v:\theta_v\geq 0}\mu_{0,v}p_v}{\sum_{v:\theta_v<0}\mu_{0,v}p_v}\sum_{v:\theta_v<0}\mu_{0,v}p_v|\theta_v|+\sum_{v:\theta_v\geq0}(1-\mu_{0,v})p_v\theta_v
\\&\geq 0\end{align*}
as required. Similarly, if $\theta\geq 0$, then
\begin{align*}
\left(1-\mu_0\right)\left(\theta^+-\theta\right)&=
\left(\frac{1-\mu_0}{\sum_{v:\theta_v\geq 0}(1-\mu_{0,v})p_v}-1\right)\sum_{v:\theta_v\geq 0}\left(1-\mu_{0,v}\right)p_v\theta_v-\sum_{v:\theta_v<0}\mu_{0,v}p_v\theta_v
\\&=\frac{\sum_{v:\theta_v<0}(1-\mu_{0,v})p_v}{\sum_{v:\theta_v\geq 0}(1-\mu_{0,v})p_v}\sum_{v:\theta_v\geq 0}\left(1-\mu_{0,v}\right)p_v\theta_v +\sum_{v:\theta_v<0}\mu_{0,v}p_v|\theta_v|
\\&\geq 0\end{align*}
which completes the proof.
\end{proof}
\pagebreak
\clearpage
\bibliographystyle{rss}

\bibliography{references}

\end{document}